\pdfoutput=1
\documentclass[sigplan,screen,sigconf]{acmart}

\setcopyright{rightsretained}
\acmDOI{}
\acmISBN{}
\acmConference[LATTE '21]{1st Workshop on Languages, Tools, and Techniques for Accelerator Design}{April 15, 2021}{Virtual, Earth}

\usepackage{tikz}
\usepackage{url}
\usepackage{graphicx}
\usetikzlibrary{backgrounds}

\begin{document}

\title[Compiler Infrastructure for Specializing Domain-Specific Memory Templates]{Compiler Infrastructure for Specializing Domain-Specific Memory Templates}

\author{Stephanie Soldavini}
\affiliation{
\institution{Politecnico di Milano}
\city{Milan}
\country{Italy}
}
\email{stephanie.soldavini@polimi.it}

\author{Christian Pilato}
\affiliation{
\institution{Politecnico di Milano}
\city{Milan}
\country{Italy}
}
\email{christian.pilato@polimi.it}

\begin{abstract}
Specialized hardware accelerators are becoming important for more and more applications. Thanks to specialization, they can achieve high performance and energy efficiency but their design is complex and time consuming. This problem is exacerbated when large amounts of data must be processed, like in modern big data and machine learning applications. The designer has not only to optimize the accelerator logic but also produce efficient memory architectures. To simplify this process, we propose a multi-level compilation flow that specializes a domain-specific memory template to match data, application, and technology requirements. 
\end{abstract}

\maketitle

\section{Introduction}

Domain-specific accelerators are one of the key solutions to continue increasing performance and efficiency beyond the end of Moore's law scaling~\cite{Esmaeilzadeh2012,dally_commacm_2020}. These accelerators use only the minimal required resources, consume less power, and compute faster than general purpose hardware~\cite{Horowitz2014}. However, the design of such components is complex~\cite{dally_commacm_2020}.

Modern big data and machine learning applications need to process huge and potentially distributed data sets with stringent requirements. Managing these data sets requires a combination of different solutions to hide the communication latency and exploit the inherent data parallelism~\cite{pilato2021everest}. Researchers proposed accelerators with local caches and private local memories for storing data on chip, while multiple channels help combine classic DRAM with non-volatile memories (NVM) for off-chip data. Memory architectures with \textbf{intelligent data transfers} can greatly optimize the systems but require specialization based on the application~\cite{mutlu2020intelligent}.

On one side, domain-specific languages like Spatial~\cite{10.1145/3192366.3192379} can abstract memory operations while  still being hardware-oriented, but they miss a complete tool-flow to port software-oriented algorithms to hardware. High-level synthesis (HLS) is a technology to automatically generate hardware modules starting from high-level descriptions~\cite{10.1109/TCAD.2015.2513673,5737854} but memory optimization is still an open problem~\cite{pilato_tcad_2017}. This line of research proposes a compiler-based approach for optimizing the accelerator memories on top of traditional HLS. The main idea is to use \textit{domain-specific annotations} to pass useful information to the compiler, transform the intermediate representations, and interface directly with modern HLS tools.

\section{High-Level Synthesis: The Present}

High-level synthesis helps raise the abstraction level and use high-level, software-like methods for hardware design. Modern high-level synthesis tools are based on state-of-the-art compilers to extract a language-agnostic intermediate representation from common software languages~\cite{10.1109/TCAD.2015.2513673}. Using compiler frontends also allows designers to apply common compiler transformations like constant propagation, dead-code elimination, and loop transformations.
For example, most HLS tools use the GCC or LLVM compilers to apply state-of-the-art compiler transformations and extract the resulting intermediate representation. In the following phase, the HLS engine determines how to distribute the operations over time (scheduling) and over the hardware resources (allocation and binding). These steps determine the hardware architecture of the \textit{controller}, which determines the evolution of the circuit in each clock cycle, and the \textit{datapath}, which contains the hardware resources and their interconnections.

Current HLS tools have strong focus on the computational aspects, while the surrounding memory architecture is adapted to merely sustain the required data accesses. 
In case of data-intensive applications, the optimizations should focus more on coordinating memory transfers and accesses, rather than on the actual computation.
To do so, compilers need to integrate, propagate, and expose more data-related information. If passed to the HLS engine, this information can help specialize the memory architecture together with the accelerators.

\section{Domain-Specific Memory Template}

\textit{Specialized architectures} are designed specifically for an accelerator, but the process is time consuming and must be done for each new design. \textit{Domain-specific architectures} are more general since the structure can be reused across multiple applications, sacrificing performance. For the memory aspects of a hardware accelerator, we propose an approach in between, using a \textbf{domain-specific template} that allows the specialization of particular components. 

The lower part of \autoref{fig:my_label} shows the proposed template. It is composed of existing memory primitives, like caches, DMA engines, prefetchers, and multi-port memories. 
Based on given area constraints, only part of the data can stay on chip, while the rest is stored in DRAM or non-volatile memories (either on the same device or remotely). 
On-chip data are stored in different memories based on the application data structures but also the type of accesses that are expected. Irregular accesses can be implemented with custom \textbf{latency-insensitive memory architectures}~\cite{10.1145/2968456.2976764}.
Data with regular accesses can be stored in fixed-latency \textbf{private local memories} (PLMs) and customized with multi-bank configurations to expose a large number of ports to the accelerator logic. Data reuse buffers can remove unnecessary data transfers.
Data accesses with a certain degree of locality can benefit from architectures featuring \textbf{caches} that are local or shared with the processor by means of a coherent protocol~\cite{10.5555/3195638.3195697,10.1145/3400302.3415753}.  
We also feature a \textbf{direct-memory access} (DMA) engine to make the data transfers more efficient and a \textbf{prefetcher} to anticipate known data transfers to hide the communication latency. These IP blocks can be augmented with \textit{special functions}, like data protection (e.g., encryption) or application-specific transformations (e.g., matrix transpose).

This template is general enough to be reused across multiple applications but it can also be specialized based on the accelerator characteristics. For instance, we can vary the number of ports on a multi-bank memory based on the specific access patterns of the application. Also, components can be removed if they are unnecessary for the application. For example, if the data resides entirely on-chip, the prefetcher can be removed or if there is only a single memory, the multi-channel controller can be simplified. We propose to use a compiler-based approach to progressively refine such template. 

\section{Specialization of the Memory Template}

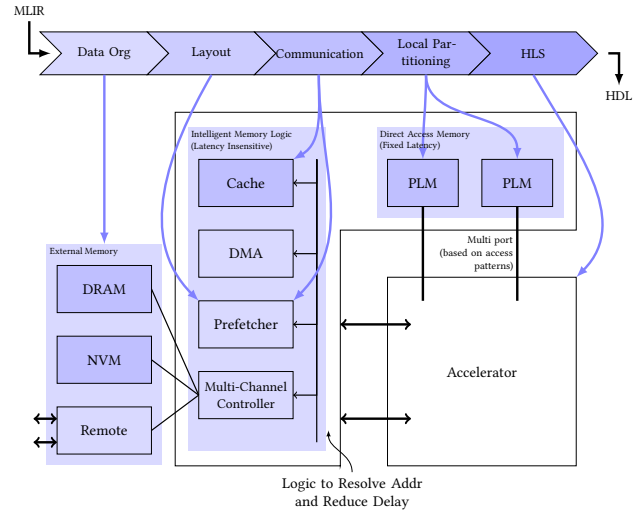
\begin{figure}[!t]
    \centering
\resizebox{\linewidth}{!}{\tikzset{
  pics/block4/.style 2 args={
      code={
    \coordinate (-n) at  (    0,  #2/2);
    \coordinate (-ne) at ( #1/2,  #2/2);
    \coordinate (-e) at  ( #1/2,     0);
    \coordinate (-se) at ( #1/2, -#2/2);
    \coordinate (-s) at  (    0, -#2/2);
    \coordinate (-sw) at (-#1/2, -#2/2);
    \coordinate (-w) at  (-#1/2,     0);
    \coordinate (-nw) at (-#1/2,  #2/2);
    \coordinate () at (0,0);
    \draw[pic actions] (-#1/2 ,#2/2) -- ++(#1,0) -- ++(0,-#2)  -- ++(-#1,0) --cycle;
    \node{ \tikzpictext};
}  } }
\tikzset{
  pics/flowarrow/.style 2 args={
      code={
    \coordinate (-n) at  (    0,  #2/2);
    \coordinate (-ne) at ( #1/2,  #2/2);
    \coordinate (-e) at  ( #1/2,     0);
    \coordinate (-se) at ( #1/2, -#2/2);
    \coordinate (-s) at  (    0, -#2/2);
    \coordinate (-sw) at (-#1/2, -#2/2);
    \coordinate (-w) at  (-#1/2,     0);
    \coordinate (-nw) at (-#1/2,  #2/2);
    \coordinate () at (0,0);
    \draw[pic actions] (-#1/2-0.25 ,#2/2) -- ++(#1,0) -- ++(0.5, -#2/2) -- ++(-0.5,-#2/2)  -- ++(-#1,0) -- ++(0.5,#2/2)--cycle;
    \node{ \tikzpictext};
}  } }

\begin{tikzpicture}
    \begin{scope}[scale=1.1, transform shape]
    \draw(0,0) coordinate (topleft) {} -- ++(4,0) coordinate (topright) {} -- ++(0,-4) coordinate (botright) {}-- ++(-4,0) coordinate (botleft) {} -- cycle;
    \draw(2,-2) node {Accelerator};
    
    \draw([shift={(-1,0)}] botleft) -| ++(-3.5,7.5)  -|  ([yshift=1cm]topright) -| cycle;

    \draw[<->,ultra thick] ([shift={(0.5,-1)}]topleft) -- ([shift={(-1,-1)}]topleft) ;
    \draw[<->,ultra thick] ([shift={(0.5,-3)}]topleft) -- ([shift={(-1,-3)}]topleft)node[below,xshift=0.5cm] (lis) {};
    \node[left of=lis,xshift=0.25cm] (addrlogic) {};
    \draw[latex-] (addrlogic) to [out=-90, in=-270] ++(0.5,-1) node[below,align=center,text width=4cm] {Logic to Resolve Addr and Reduce Delay};

    \draw pic[left of=topleft, shift={(-2,2)},fill=blue!25] (a) {block4={2}{1}};
    \node[align=center,text width=2cm] at (a) {Cache};
    \draw[->,thick] (-1.5,2.5) |- (a-e)node[above left] {};
    
    \draw pic[below of=a,yshift=-0.5cm] (b) {block4={2}{1}};
    \node[align=center,text width=2cm] at (b) {DMA};
    \draw[->,thick] (-1.5,2.5) |- (b-e)node[above left] {};
    
    \draw pic[below of=b,yshift=-0.5cm] (c) {block4={2}{1}};
    \node[align=center,text width=2cm] at (c) {Prefetcher};
    \draw[->,thick] (-1.5,2.5) |- (c-e)node[above left] {};
    
    \draw pic[below of=c,yshift=-0.5cm] (ctrl) {block4={2}{1}};
    \node[align=center,text width=2.5cm,scale=0.9] at (ctrl) {Multi-Channel Controller};
    \draw[->,thick] (-1.5,2.5) |- (ctrl-e);

    \draw[-,thick] (-1.5,2.5) -- ++(0,-6);
    
    \draw pic[left of=ctrl,xshift=-2cm,yshift=2.25cm,fill=blue!25] (dram) {block4={2}{1}};
    \node[align=center,text width=2cm] at (dram) {DRAM};
    \draw[-,thick] (ctrl-w) -- (dram-e) node[above left] {};

    \draw pic[below of=dram,yshift=-0.5cm,fill=blue!25] (nvm) {block4={2}{1}};
    \node[align=center,text width=2cm] at (nvm) {NVM};
    \draw[-,thick] (ctrl-w) -- (nvm-e) node[above left] {};

    \draw pic[below of=nvm,yshift=-0.5cm] (remote) {block4={2}{1}};
    \node[align=center,text width=2cm] at (remote) {Remote};
    \draw[-,thick] (ctrl-w) -- (remote-e) node[above left] {};
    \draw[<->,ultra thick] ([yshift=0.25cm]remote-w) -- ++(-0.5,0);
    \draw[<->,ultra thick] ([yshift=-0.25cm]remote-w) -- ++(-0.5,0);

    \draw pic[above of=topleft,shift={(0.75,1)},fill=blue!25] (d) {block4={1.5}{1}};
    \node[align=center,text width=2cm] at (d) {PLM};
    \draw[-,ultra thick] (d-s) -- ++(0,-2);

    \draw pic[above of=topleft,shift={(2.75,1)},fill=blue!25] (e) {block4={1.5}{1}};
    \node[align=center,text width=2cm] at (e) {PLM};
    \draw[-,ultra thick] (e-s) -- ++(0,-2) node[left, align=right,yshift=1cm,scale=0.7,text width = 5cm] (da) {Multi port\\(based on access\\patterns)};
    \end{scope}


    \draw pic[above of=dram,yshift=4.5cm,fill=blue!15] (dataorg) {flowarrow={2.5}{1}};
    \node[align=center,text width=2cm] at (dataorg) {Data Org};
    \draw[-latex,ultra thick, blue!50] (dataorg-s) -- ([yshift=0.5cm]dram-n);
    
    \draw pic[right of=dataorg,xshift=1.5cm,fill=blue!20] (layout) {flowarrow={2.5}{1}};
    \node[align=center,text width=2cm] at (layout) {Layout};
    \draw[-latex,ultra thick, blue!50] (layout-s) to [out=240, in=135] (c-nw);
    
    \draw pic[right of=layout,xshift=1.5cm,fill=blue!25] (comp) {flowarrow={2.5}{1}};
    \node[align=center,text width=2cm] at (comp) {Communication};
    \draw[-latex,ultra thick, blue!50] (comp-s) to [out=270, in=30] (a-ne);
    \draw[-latex,ultra thick, blue!50] (comp-s) to [out=270, in=45] (c-ne);

    \draw pic[right of=comp,xshift=1.5cm,fill=blue!30] (locpart) {flowarrow={2.5}{1}};
    \node[align=center,text width=2cm] at (locpart) {Local Partitioning};
    \draw[-latex,ultra thick, blue!50] (locpart-s) to [out=270, in=90] (d-n);
    \draw[-latex,ultra thick, blue!50] (locpart-s) to [out=270, in=90] (e-n);

    \draw pic[right of=locpart,xshift=1.5cm,fill=blue!35] (hls) {flowarrow={2.5}{1}};
    \node[align=center,text width=2cm] at (hls) {HLS};
    \draw[-latex,ultra thick, blue!50] (hls-s) to [out=300, in=45] (topright);

    \draw[latex-,ultra thick] (dataorg-w) -| ++(-0.5,0.75) node[above] {MLIR};
    \draw[-latex,ultra thick] ([xshift=0.5cm]hls-e) -| ++(0.25,-0.75) node[below] {HDL};
    
    \begin{scope}[on background layer]
        \fill[fill=blue!15] ([shift={(-0.25,0.5)}]dram-nw) node[below right, scale=0.7] {{External Memory}} -| ([shift={(0.25,-0.25)}]remote-se) -| cycle;
        \fill[fill=blue!15] ([shift={(-0.25,0.75)}]a-nw) node (iml) {}-| ([shift={(0.75,-0.75)}]ctrl-se) -| cycle;
        \fill[fill=blue!15] ([shift={(-0.25,0.75)}]d-nw) node (dam) {} -| ([shift={(0.25,-0.25)}]e-se) -| cycle;
    \end{scope}
    \node [below right,scale=0.7,text width=6cm]at (iml) {{Intelligent Memory Logic}\\(Latency Insensitive)};
    \node[below right, scale=0.7,text width=6cm] at (dam) {{Direct Access Memory}\\(Fixed Latency)};
    
\end{tikzpicture}}
    \vspace{-20pt}\caption{Multi-level compilation flow for the specialization of domain-specific memory architectures.}\vspace{-4pt}
    \label{fig:my_label}
\end{figure}

To achieve better performance and reduce costs, designers can specialize the memory template based on the given accelerator. For this, our approach is based on the idea of \textit{platform-based design}~\cite{10.1109/54.970421}, where the memory template is refined in different stages, starting from the general organization of the data in memory to the actual interaction with the actual accelerator.
The upper part of \autoref{fig:my_label} shows our compiler-based customization flow.  

\vspace{4pt}\noindent{\bf Intermediate Representation. }
The compiler infrastructure will need to include more hardware-related information. 
We target novel multi-level representations, like MLIR~\cite{lattner2020mlir}, to include more hardware-related information early in the compilation flow to make progressive refinements of the architecture at proper levels of abstraction. 
A novel flow is required because existing approaches are not fully compatible with HLS. CIRCT~\cite{circt} proposes MLIR extensions for low-level hardware synthesis (below the HLS level). Calyx~\cite{nigam2021compiler} follows, instead, a different approach with a novel IR and associated compiler. SODA~\cite{9256693} proposes a MLIR-based synthesis framework for machine learning accelerators with more focus on the computational aspects.

\vspace{4pt}\noindent{\bf Compilation Flow. }
We extend the LLVM-MLIR compilation flow with additional passes to include memory-related information and transform the IR accordingly.
Our passes include solutions to define the data layout, size the physical memories (both caches and PLMs), optimize the access patterns, and create multi-port PLMs for fast access. Currently, we use custom generators like Mnemosyne\footnote{http://github.com/chrpilat/mnemosyne} to derive the HDL descriptions from such information. We will also investigate the possibility to interface directly with MLIR formats for hardware, like CIRCT.

The customization flow shown at the top of \autoref{fig:my_label} would proceed as follows: At the highest abstraction level, the \textit{data organization} phase analyzes the data representations to determine the coarse memory structure, i.e. deciding which data are stored off-chip or on-chip. The next step, the \textit{layout} phase, reorganizes the computation to better exploit local memories (either caches or PLMs). Then, in the \textit{communication} phase, the prefetcher is configured to hide transfer latency based on the data access patterns. After this, the \textit{local partitioning} phase determines the multi-bank PLM architecture, also sharing physical memories for data with disjoint lifetimes~\cite{pilato_tcad_2017}. Finally, the \textit{HLS} phase generates the computation part of the component with traditional HLS, producing the complete syntesizable description of the accelerator.

\vspace{4pt}\noindent{\bf Accelerator Logic HLS. }
With our approach, the accelerator is designed only at the end of the flow according to the resulting memory organization. The accelerator features state-of-the-art solutions for memory management (e.g., dynamic address resolution~\cite{Pilato2011,Pilato2013}). The accelerator is mostly unaware of the data organization and layout since the IR has been already updated based on the memory transformations. It is only optimized to efficiently access the data with fixed or unbounded latency. This part can leverage existing HLS tools that start from low-level intermediate representations. For example, the final LLVM IR representation can be directly interfaced with  the Xilinx Vitis HLS front-end\footnote{https://github.com/Xilinx/HLS}.

\section{Conclusion}

We described a novel approach for specializing domain-specific memory templates during the compilation flow and \textit{before} high-level synthesis of the accelerator logic. Starting from a high-level memory template, we apply a multi-level compilation flow based on MLIR that progressively refines the memory architecture and then interfaces with commercial HLS tools. Our approach borrows idea from platform-based design, trading off flexibility and specialization based on specific needs of the designers.

\section*{Acknoledgements}

This project is partially funded by the EU Horizon 2020 Programme under grant agreement No 957269 (EVEREST).

\newpage 

\balance
\bibliographystyle{ACM-Reference-Format}
\bibliography{main}

\end{document}